\documentclass[a4paper,11pt]{article}
\usepackage{pos}
\usepackage{multirow}
\usepackage{graphicx}
 \usepackage{url}
\usepackage{appendix}
\usepackage{color,soul}
\usepackage{amsmath}
\usepackage{float}
\usepackage{subfig}

\title{Migration of CMSWEB Cluster at CERN to Kubernetes}

\author*[a,b]{Muhammad Imran}
\author[c]{Valentin Kuznetsov}
 \author[a]{Lina Marcella}
 \author[a]{Katarzyna Maria Dziedziniewicz-Wojcik} 
 \author[a]{Andreas Pfeiffer}
 \author[a]{Panos Paparrigopoulos}

\affiliation[a]{CERN,\\
  Meyrin, Geneva, Switzerland}
\affiliation[b]{National Centre for Physics,\\
  Islamabad, Pakistan}

\affiliation[c]{Cornell University,\\
 New York, USA}

\emailAdd{muhammad.imran@cern.ch}
\emailAdd{vkuznet@protonmail.com}
\emailAdd{becerraji87@gmail.com}
\emailAdd{katarzyna.maria.dziedziniewicz@cern.ch}
\emailAdd{andreas.pfeiffer@cern.ch}
\emailAdd{panos.paparrigopoulos@cern.ch}

\abstract{The CMS experiment heavily relies on the CMSWEB cluster to host critical services for its operational needs. The cluster is deployed on virtual machines (VMs) from the CERN OpenStack cloud and is manually maintained by operators and developers. The release cycle is composed of several steps, from building RPMs, their deployment to perform validation, and integration tests. To enhance the sustainability of the CMSWEB cluster, CMS decided to migrate its cluster to a containerized solution such as Docker, orchestrated with Kubernetes (k8s). This allows us to significantly reduce the release upgrade cycle, follow the end-to-end deployment procedure, and reduce operational cost. This paper gives an overview of the current CMSWEB cluster and its issues. We describe the new architecture of the CMSWEB cluster in Kubernetes. We also provide a comparison of VM and Kubernetes deployment approaches and report on lessons learned during the migration process.
}

\FullConference{%
  40th International Conference on High Energy physics - ICHEP2020\\
  July 28 - August 6, 2020\\
  Prague, Czech Republic (virtual meeting)
}

\begin{document}
\maketitle

\section{Introduction\label{intro_sect}}
The Compact Muon Solenoid (CMS) is a general-purpose detector at the Large Hadron Collider (LHC) at CERN, Geneva, Switzerland \cite{cms}. 
The CMS experiment runs hundreds of thousands of jobs daily on its distributed computing system to simulate, reconstruct and analyse the data taken during collision runs. A dedicated cluster ("CMSWEB") is used to host essential CMS central services which are responsible for the CMS data management, data discovery, and various data bookkeeping tasks. The cluster is based on virtual machines (VMs) on the CERN OpenStack cloud infrastructure. Each service is managed by its own development team. Due to the complexity of the heterogeneous environment, different schedules of development teams, only monthly release cycles can be afforded. Each upgrade cycle includes: the build of RPMs from source code, including all dependency chains, the cross-validation of all software components, and the validation of the correct interactions of all services. This typically requires a lot of communication between development teams and operators, as well as coordination of various efforts in a coherent manner. 

To enhance the sustainability of CMSWEB, CMS decided to migrate to a containerized solution based on Docker and orchestrated with Kubernetes (``k8s''), a de facto industry standard for managing containers \cite{k8s}.  Recently, an instance of the testbed CMSWEB cluster was successfully migrated to Kubernetes. With the containerized approach, developers will not have to ask the operators to deploy their services, they  can deploy new versions of their services in a few seconds. This significantly reduces the release upgrade cycle, follows the end-to-end deployment procedures, and reduces operational cost. 

In this paper, an overview of the current CMSWEB cluster and the issues  faced in this cluster is given. The new architecture of the CMSWEB cluster orchestrated by Kubernetes is described. Furthermore, various issues found during the implementation in Kubernetes are discussed. We also provide a comparison of VM and Kubernetes deployment approaches, and report on lessons learned during the migration process.

The remainder of this paper is organized as follows. Section \ref{Architectur} describes the current architecture of CMSWEB cluster. Section \ref{Deployment} presents the proposed architecture of CMSWEB in Kubernetes. Section \ref{Performance} presents the performance analysis. We discuss our plans for the future in Section \ref{FutureWork}. We conclude in Section \ref{conclusion_sect}.  

\section{Architecture of CMSWEB\label{Architectur}}
The CMSWEB cluster consists of 21 services maintained by CMS operators. 
The CMSWEB cluster allows: independent development and evolution of underlying services; simplified integration and regression testing when rolling out new service versions; and  building external services that integrate information from several sources in a clean manner.

\begin{figure*}[!t]
\centering
\includegraphics[scale=0.8]{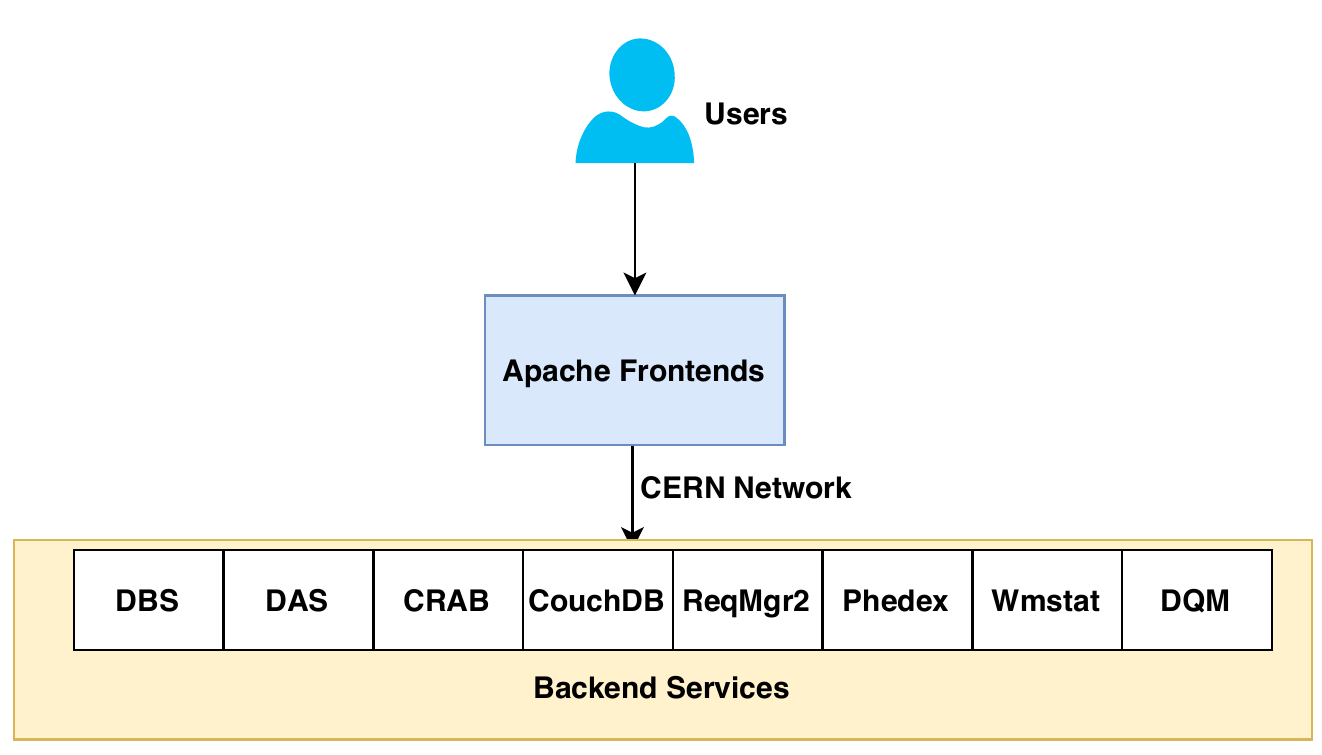}
\caption{The CMSWEB VM Cluster Architecture 
}
\label{vm_cluster}
\end{figure*}

The architecture of the CMSWEB VM cluster is shown in Figure \ref{vm_cluster}. It has two layers of services i.e., frontend and backend services. The frontend service is based on Apache which performs authentication using certificates and redirects requests to the requested backend service, while the backend services perform their relevant tasks. The frontend service has redirect rules to forward the requests to the relevant VM node running the backend service. 


\begin{figure*}[!t]
\centering
\includegraphics[scale=0.7]{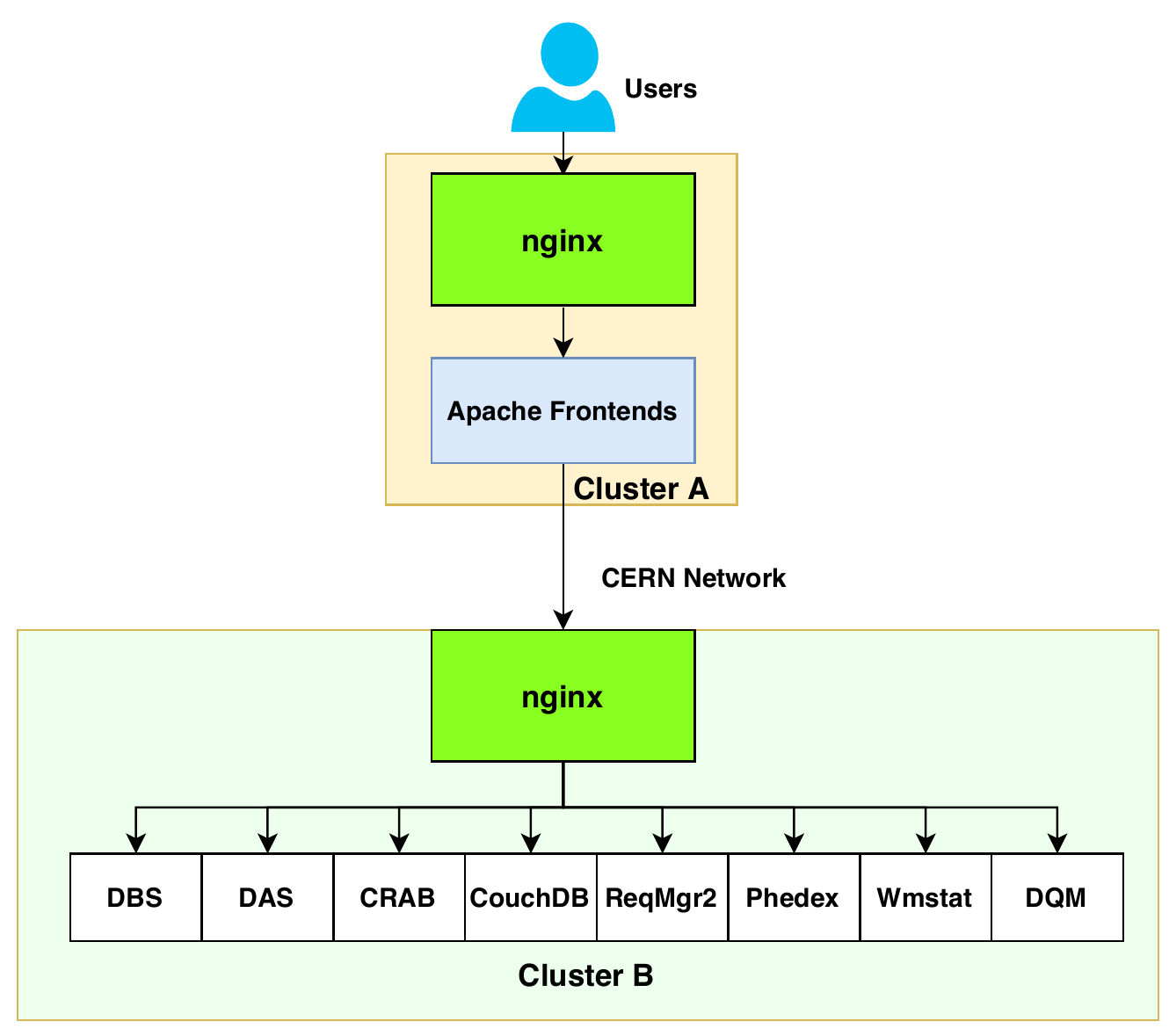}
\caption{The CMSWEB Kubernetes cluster architecture
}
\label{cmsweb_architecture_k8s}
\end{figure*}

\section{Deployment of CMSWEB cluster to Kubernetes\label{Deployment}}


The proposed architecture of CMSWEB in Kubernetes is shown in Figure \ref{cmsweb_architecture_k8s}. It has the two components: frontend cluster (cluster A); and backend cluster (cluster B). The frontend cluster contains the CMSWEB Apache frontend behind the Nginx Kubernetes ingress controller (server). The backend cluster contains all CMSWEB back-end services behind its ingress controller. The frontend cluster ingress controller provides transport layer security (TLS) passthrough capabilities to pass client's requests (with certificates) to the Apache frontend. The Apache frontend performs CMSWEB authentication and redirects the request to the backend cluster. On the backend cluster, the ingress controller has basic redirect rules to the appropriate services and only allows requests from the frontend cluster.


To host services in Kubernetes, we need container images of all services, which are then deployed in Kubernetes. The Docker images are created using RPMs of the current VM cluster, and are kept in a central repository which is available at \cite{cmsweb_docker_repo}. 
Similarly to the \emph{docker} area in the repository, there is also a \emph{kubernetes} area for deploying those images in Kubernetes. 
A central \emph{cmsweb} directory in the repository contains sub-directories for all CMSWEB namespaces \cite{cmsweb_k8s_repo}. 

\begin{figure*}[!t]
\centering
\includegraphics[scale=0.45]{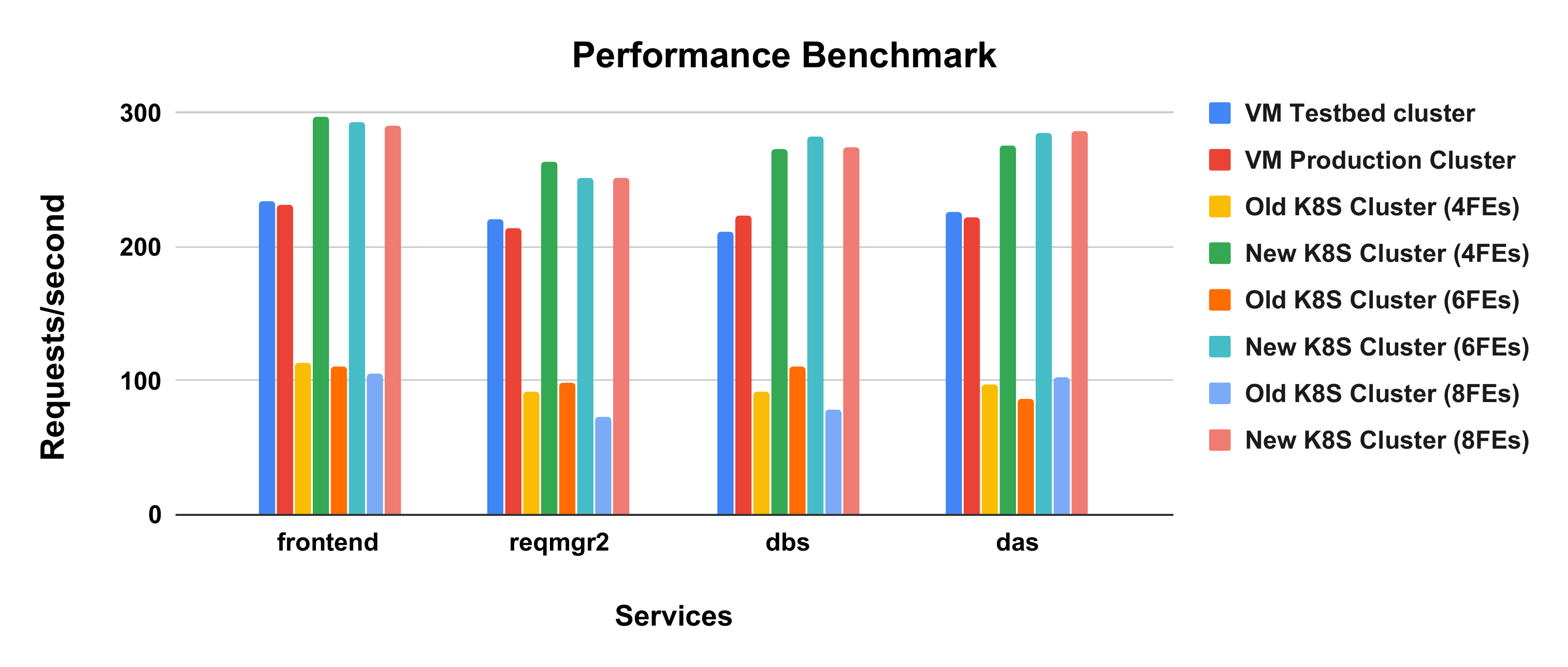}
\caption{The performance benchmark results  in the form of requests/second (y-axis) of some of the commonly used CMSWEB services (x-axis) for various clusters (i.e. VM production cluster, VM testbed cluster, old Kubernetes cluster and new Kubernetes cluster using a different number of replicas.). The $100$ tests were performed for each configuration and results show average values from these tests. 
}
\label{result1}
\end{figure*}
\section{Performance Analysis\label{Performance}}
We used the \emph{hey} tool \cite{hey} to evaluate the performance of the VM based clusters (both testbed and production) and the new testbed clusters in Kubernetes. The initial Kubernetes cluster which used version 1.15, showed a severe network degradation caused by a faulty network driver (``flannel''). This issue was fixed in  Kubernetes version 1.17, which uses the calico network driver. In the following comparison, we label the clusters using Kubernetes version 1.15 as ``old cluster'', and the ones using version 1.17 as ``new cluster''.









To test the performance, 
a configuration with $n=10$ and $c=5$ was chosen, where $n$ is the number of requests to run and $c$ is the number of workers to run concurrently. 

The performance results are shown in Figure \ref{result1}: some services of CMSWEB are shown in the x-axis while the resulting requests/second are shown on the y-axis. A comparative analysis of the VM production cluster, VM testbed cluster, old Kubernetes cluster, and new Kubernetes cluster was performed. The old and new Kubernetes clusters were studied with different numbers of replicas of frontend i.e. (4,6,8) to see the impact of replicas on the overall performance. It can be seen that the new Kubernetes cluster performs much better as compared to the production, testbed and old Kubernetes cluster for all services benchmarked.

\subsection{Issues Faced During Migration\label{issues_sect}}
During this migration process various issues were encountered, which are categorized as either infrastructure or service issues
\subsubsection{Infrastructure Issues\label{infrastructures_issues_sect}}
\textbf{Network Degradation:}
A major issue of network degradation was identified by us and other groups, which was related to the infrastructure at CERN, 
Further detail of the performance benchmark is available in Section \ref{Performance}. 

\textbf{Cluster Creation Issues:}
With the new Kubernetes version 1.17, cluster creation failed due to timeout. This issue was related to storage volumes on servers which were in different availability zones and had higher latencies, causing timeouts. 

\textbf{Ceph Mount Issues:}
A problem with Ceph mounts showed up after migrating clusters to the new Kubernetes version. This issue was caused by a version of the cloud client which CERN has provided for the management and creation of clusters in the CERN network, which was not compatible with the new version of Kubernetes. 

\textbf{Permission Mount Issues:}
Permission issues of the mount point in /etc/grid-security in the new Kubernetes version were found to be related to the security context of an unconfigured pod. A security context defines privilege and access control settings for a Pod or Container 

\textbf{Nginx-ingress controller Issues:}
A problem with nginx-ingress controller was encountered when we performed stress tests on the K8s cluster. It was related to the low value of file descriptors in the nginx-ingress controller. During stress tests many requests were failed because of that. 

These issues were fixed with the help of CERN IT.


\subsubsection{Service Issues\label{service_issues_sect}}
\textbf{CouchDB:}
We noticed that CouchDB crashed in the Kubernetes cluster. Handling things like databases, availability to other layers of the application, and redundancy for a database can have very specific requirements. That makes it challenging to run a database in a distributed environment. 
It was therefore decided to keep this service in the VM cluster and all CouchDB requests are redirected to the VM cluster.

\textbf{PhEDEx:}
The PhEDEx service is one of the legacy application we are required to support during the transition to Kubernetes. 
So, it was decided to not spend time on porting it to new infrastructure and keep it in the dedicated VMs.

\textbf{DBS:}
Initially, in the Kubernetes cluster, the same accounts as the ones of the VM based clusters were used, this caused issues. 
In order to avoid potential overwrite of data in production DBS DB instances, we were asked by DBS team to use separate accounts for DBS in Kubernetes cluster.


\subsection{Lessons Learned\label{lessons_learned_sect}}

The new Kubernetes infrastructure automates the procedure of service deployment as the developers are able to deploy their services directly in the Kubernetes cluster without needing input from the CMSWEB operator. 

The current VM cluster lacks autoscaling feature; every service is deployed on the particular VM nodes and the resources are assigned on the VM level instead of the service level. When the services are overloaded, they often become unresponsive, then the CMSWEB operator manually interferes and restarts individual services. The auto-scaling feature of Kubernetes, however, automatically manages the resources based on the workload. 

Manifest files in Kubernetes require verification. A small typo and wrong indentation leads to failures. 

\section{Future work\label{FutureWork}}
We plan to work on the following items:

{\bf Custom auto-scalers}, the default settings of K8s only provides auto-scaling based on CPU and RAM usage of the pods. Since we run several applications within a pod and monitor their usage via the Prometheus service, we can take advantage of using service metrics for auto-scalers and perform dynamic tuning of services based on those metrics.

{\bf Service-mesh deployment}, the service-mesh provides plenty of benefits to Kubernetes, including traffic encryption within the cluster, traffic routing between different releases, canary deployment and rolling release cycles. We would like to bring this functionality to our infrastructure via the Istio \cite{istio}.
\section{Conclusions\label{conclusion_sect}}


In this paper, we give an overview of the current CMSWEB cluster and the issues which we faced in the cluster. We describe the new architecture of the CMSWEB cluster in Kubernetes and explain the implementation strategy of the proposed architecture in Kubernetes. Furthermore, we describe various issues that we faced during the implementation in Kubernetes. We also provide a comparison of VM and Kubernetes deployment approaches, emphasizing the pros and cons of the new architecture and report on lessons learned during the migration process. The new cluster of CMSWEB in Kubernetes enhances sustainability and reduces the operational cost of CMSWEB. 

\bibliographystyle{plain}
\bibliography{References}

\begin{thebibliography}{1}

\bibitem{cms}
{CMS experiment at CERN}.
\newblock Website: \url{https://home.cern/science/experiments/cms}.
\newblock last accessed: 04.13.2020.

\bibitem{cmsweb_docker_repo}
Cms repository for docker.
\newblock Website:
  \url{https://github.com/dmwm/CMSKubernetes/tree/master/docker}.
\newblock last accessed: 04.04.2020.

\bibitem{cmsweb_k8s_repo}
Cmsweb kubernetes repository.
\newblock Website:
  \url{https://github.com/dmwm/CMSKubernetes/tree/master/kubernetes}.
\newblock last accessed: 04.04.2020.

\bibitem{hey}
Hey tool.
\newblock Website: \url{https://github.com/rakyll/hey}.
\newblock last accessed: 04.04.2020.

\bibitem{istio}
Istio middleware.
\newblock Website: \url{https://istio.io/}.
\newblock last accessed: 04.16.2020.

\bibitem{k8s}
Kubernetes.
\newblock Website: \url{https://kubernetes.io/docs/home/}.
\newblock last accessed: 04.04.2020.

\end{thebibliography}


\end{document}